\documentclass[conference]{IEEEtran}

\usepackage{ amsmath, amssymb, multicol, graphicx}
\usepackage{amsmath,amsfonts}
\usepackage{algorithmic}
\usepackage{textcomp}
\usepackage{xcolor}
\usepackage{flushend}

\begin{document}

\title{Water Purification Via Plasma}

\author{{Angel Gonzalez-Lizardo}, {Gretchen Morales}, {Jairo Rond\'on}, {Jennifer Santiago}, {Adrián Landrón}, {Xavier Peña}}
\thanks{Polytechnic. University of Puerto Rico}

\author{\IEEEauthorblockN{Angel Gonzalez-Lizardo\IEEEauthorrefmark{1}, Gretchen Morales\IEEEauthorrefmark{2}, Jairo Rond\'on\IEEEauthorrefmark{3}, Jennifer Santiago\IEEEauthorrefmark{4}, Adrián Landrón\IEEEauthorrefmark{5}, Xavier Peña\IEEEauthorrefmark{6}}%
\bigskip
\IEEEauthorblockA{\IEEEauthorrefmark{1}Polytechnic University of Puerto Rico,\\
 agonzalez@pupr.edu}
\IEEEauthorblockA{\IEEEauthorrefmark{2}NASA--Glenn Research Center,\\
 gmoralesvalle@outlook.com}
\IEEEauthorblockA{\IEEEauthorrefmark{3}Polytechnic University of Puerto Rico,\\
 jrondon@pupr.edu}
\IEEEauthorblockA{\IEEEauthorrefmark{4} Boeing Aviation \& Aerospace,\\
 jenny.stgo@hotmail.com}
\IEEEauthorblockA{\IEEEauthorrefmark{5} Pratt \& Whitney, Controls Dept, Puerto Rico.\\
 adrian.landron@hotmail.com}
\IEEEauthorblockA{\IEEEauthorrefmark{6}JC Automation, Execution Systems, Puerto Rico.\\
 xavierpena3@hotmail.com}}

\maketitle

\begin{abstract}
Water purification via plasma is considered a healthy, effective alternative to traditional water purification systems due to the lack of harmful chemical additives that traditional purification systems use. A reactor capable of eliminating bacteria colonies from contaminated water using plasma discharge was designed and tested. Materials and reactor parameters are discussed in this paper. It was confirmed that glow discharge plasma in water creates oxidants, eliminating bacteria colonies. 
\end{abstract}

\section{Introduction}

Many natural occurring plasmas, such as the surface regions of the sun, exhibit distinctively plasma-dynamical phenomena arising from the effects of electric and magnetic forces. The word plasma designates a fully or partially ionized gas consisting of electrons and ions. When enough energy or heat is applied to a gas, the atoms collide with each other and knock their electrons off in the process and plasma, known as the fourth state of matter, is formed \cite{nishikawa2000plasma}. Consequently, the ionized gas gain unique chemical and electrical properties. Moreover, in most materials the dynamics of motion are determined by forces between near-neighbor regions of the material. In a plasma, charge separation between ions and electrons gives rise to electric fields and charged particle flows give rise to currents and magnetic fields, which results in “action at a distance” \cite{goldston2020introduction}. Unlike the other states of matter (solid, liquid, gas), plasma mostly does not exist on Earth’s surface under normal conditions and must be artificially generated from neutral gases. Plasma can be generated by other means of energy than heat, since most of containers cannot withstand the amount of heat a plasma needs to be ionized. Commonly, a small amount of gas is heated and ionized by driving an electric current through it, or by shinning radio waves into it. Generally, these means of plasma formation give energy to free electrons in the plasma directly. Then the electron-atom collision liberates more electrons, and the process continues until the desired degree of ionization is reached \cite{Barillas_2015}.

A more detailed description of the most commonly known processes of photo ionization and electric discharge in gases is presented next, along with its chemical/electrical properties and explaining. In the photo ionization process, ionization occurs by absorption of incident photons whose energy is equal to, or greater than, the ionization potential of the absorbing atom. The excess energy of the photon is transformed into kinetic energy of the electron-ion pair formed. Also, the ionization can be produced by x-rays or gamma rays, which have much smaller wavelength. In an electrical discharge, an electric field is applied across the ionized gas, which accelerates the free electrons to energy values sufficiently high to ionize other atoms by collisions. One characteristic of this process that exceeds over the photo-ionization process is the applied electric field, which transfers energy much more efficiently to the lightweight electrons than to the relatively heavyweight ions. Therefore, the electron temperature in electrical discharges is usually higher than the ion temperature since the transfer of thermal energy from the electrons to the heavier particles is much slower. When the ionizing source is turned off, the ionization decreases gradually due to recombination until it reaches an equilibrium value consistent with the temperature of the medium. In practice, the recombination process usually occurs so fast that the plasma completely disappears in a small fraction of a second \cite{osti_1043912}.

\section{Plasma Discharges}

Plasma Discharges could be used as a replacement for chlorination \cite{defolleto} in traditional wastewater treatments, given that long term use of chlorine has proven to be harmful for the human body. Plasma can be an alternative for industries which require to treat water before disposing it into natural resources, or as a purifying method for areas with limited access to clean water. Three types of plasma discharges are often used in prototypes created for this purpose: dielectric barrier discharge, glow discharge and pulsed corona discharge \cite{zeghioud2020review}.

\subsection{Dielectric Barrier Discharge}

Dielectric barrier discharges (DBDs) refer to a type of self-pulsing plasma that occurs between insulated surfaces. Initially utilized for ozone production in 1857, the application of DBDs has expanded to include areas such as water purification. These discharges are enabled by a configuration that allows for a nonthermal plasma generation at atmospheric pressure. Various materials, including quartz, glass, plastic, and ceramics, serve as the dielectric barriers. The generation of plasma at atmospheric conditions requires high voltage across a small gap ranging from 0.1 to 10 mm. Figure 3.1 illustrates various configurations for producing DBDs. Notably, DBDs operate silently, distinguishing them from corona discharges, which are typically accompanied by noise. Predominantly, DBDs are produced using alternating current (AC) voltage in the kilohertz range, as documented in studies and experiments.

\subsection{Pulsed Corona Discharge}

\begin{figure}
  \centering
  \includegraphics[width=.8\columnwidth]{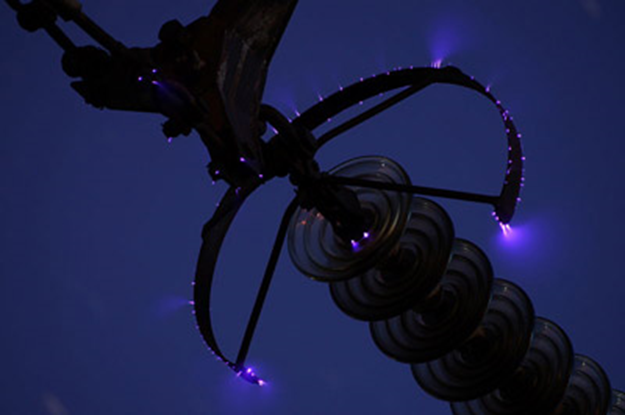}
  \caption{Corona discharge on corona ring of 500 kV overhead power line}\label{F3-2}
\end{figure}

Corona discharge (Figure \ref{F3-2}) is an electrical phenomenon that occurs when an electrical field overloads in an ionized medium, creating a visible and audible effect. It manifests as a hissing noise, a violet glow, and a lightning bolt-like appearance, with the noise intensifying as the output voltage increases. This process can generate ozone, result in power loss, and interfere with nearby radio signals. The glow is attributed to the recombination of atoms, releasing photons.
The phenomenon is produced through an electric arc, which is non-thermal and non-equilibrium in nature. It occurs when a potential difference is applied between two electrodes, generating an electric field around them, especially intense at the surface. Free electrons near the electrodes gain velocity and energy, leading to high velocity collisions with neutral molecules as the applied voltage and electric field increase. This results in ionization, where air becomes conductive and allows an arc to form. The distance between electrodes and the applied voltage are crucial for corona discharge, with closer electrodes and higher voltages facilitating the process. Corona discharges can be positive or negative, determined by the electrode's polarity, with positive discharges having lower electron density but higher energy concentration, and negative discharges appearing larger due to the spread of electrons.

\section{Glow Discharge}

Electric glow discharge, a type of plasma characterized by its glow, occurs when voltage is applied across electrodes set at a specific distance \cite{corke2010dielectric}. This setup energizes free electrons, which then accelerate towards the anode, achieving high velocities due to their low mass. This rapid movement and subsequent collisions with molecules in the surrounding medium (air, gas, or water) produce photons, causing the observable luminosity. As voltage increases, the medium becomes ionized, filled with positive ions and electrons. Positive ions move towards the cathode, and electrons towards the anode, driven by the electric potential. This process continues as long as voltage is applied.

Ions interacting with neutral particles transfer kinetic energy, leading to further collisions and sometimes causing sputtering at the cathode, where free atoms are ejected into the discharge process. These atoms, upon colliding and becoming excited, release energy quickly, often as radiation in the form of photons, contributing to the discharge's luminosity. The emitted photon's wavelength can reveal the chemical elements present and their concentrations.

This ionization mechanism is central to producing electrolysis in water, a process that alters water's composition by passing a current through it, generating antioxidant agents that can eliminate bacteria.

\begin{figure}
  \centering
  \includegraphics[width=\columnwidth]{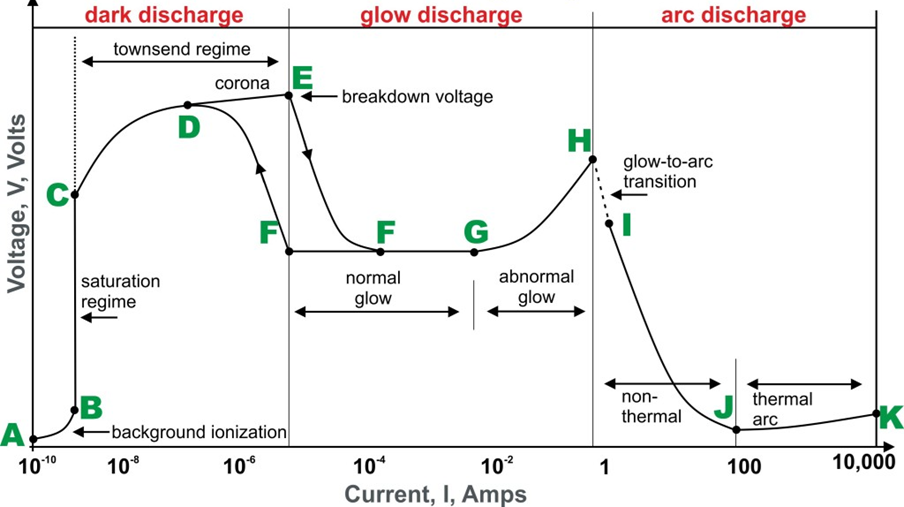}
  \caption{Electric discharge regimes for glow discharge}\label{F3-4}
\end{figure}

\section{Reactor}
\subsection{Vessel}
\begin{figure}
  \centering
  \includegraphics[width=.6\columnwidth]{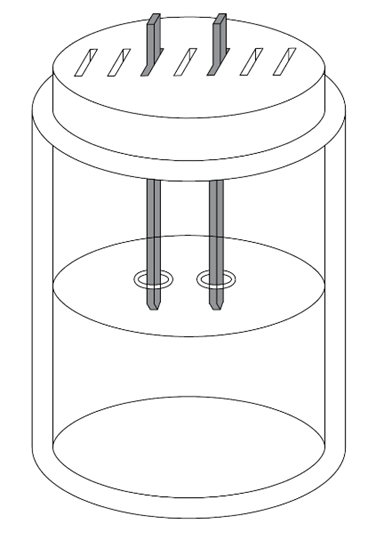}
  \caption{Point-to-point reactor}\label{F1}
\end{figure}

A Mason jar \cite{EM2020} was used as the vessel of the reactor. The main characteristic needed for the container was to be able to see through it. Mason jars were the first container to bleach glass, allowing users to see the contents. The jar was also chosen due to the material’s resistance to temperature change. Figure \ref{F1} presents the reactor with its electrodes. The reactor was 6.55 in (166.4 mm) tall with a diameter of 3.24 in (82.3 mm).

\subsection{Electrodes}

An electrode is a conductor that makes contact with a non-metallic part of a circuit, typically made from conductive metals like steel, platinum, titanium, tungsten, and copper. In this case, stainless steel was chosen for the reactor's electrodes, favored for its conductivity, durability, heat resistance, and the capability to be shaped in the Polytechnic University of Puerto Rico’s (PUPR) Materials Laboratory. The dimensions of both the anode and cathode were 161.9 mm in length and 9.5 mm in width. The gap between electrodes is crucial for plasma discharge, with a larger distance necessitating higher voltage for discharge initiation. Consequently, a range of 10 mm to 30 mm was selected for the electrode spacing.

\subsection{Power Supply}
The power supply used for the experiment was a Spellman SL 1200 \cite{SPELLMANSL1200}. This power supply outputs a voltage range of 0-1.5kV and a current range of 0-1A. A power supply capable of providing more watts than the project needed was chosen to permit the system to operate a determined amount of time without suffering any damages.

\subsection{Applied Voltage and Current}
\begin{figure}
  \centering
  \includegraphics[width=.8\columnwidth]{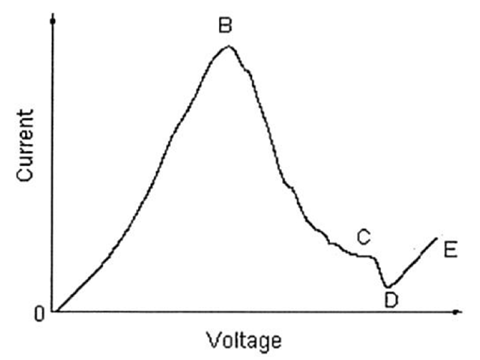}
  \caption{I-V characteristic during glow discharge electrolysis process}\label{F2}
\end{figure}

A current versus voltage (I-V) graph was used as reference, To ensure the expected discharge was being produced,. Figure \ref{F2} depicts the common I-V behavior during the glow discharge. The area between 0 and B is known as the ohmic region, in which the voltage and current have a linear relationship. As the voltage increases, so does the current. In this region, bubbles can be observed around the electrodes, and electrolysis starts to occur. When the voltage reaches what is known as its breakdown value (point B), the current will begin to decrease as the voltage increases. Here, abnormal fluctuations in the current values are observed. This is due to the formation of a layer of gas surrounding the anode. Sometimes, the area between C and D can form a horizontal line when the anode is submerged underwater. This region is known as the transition region (B to D). Once point D is reached, the current will start increasing along with the voltage, and luminosity will be more visible as the voltage increases \cite{Wang2012ApplicationOG}.

\section{Water treatment with electrolysis}

Plasma discharges in water facilitate the process of electrolysis, which is the alteration of a medium's chemical composition through the passage of an electrical current. This current prompts the substance to either gain or lose electrons, leading to oxidation or reduction reactions. Such reactions can be conducted within a reactor using two conductive metal electrodes placed at a specific distance from each other, either immersed in or adjacent to an ion-charged solution. The ion exchange between the electrodes and the solution generates free hydroxyl ($\bullet OH$) and hydrogen ($\bullet H$) radicals. The $\bullet OH$ radicals, which have a brief existence, tend to quickly recombine to form hydrogen peroxide ($H_2O_2$). Both $\bullet OH$ radicals and $H_2O_2$ act as oxidizing agents, meaning that their presence, especially during a glow discharge in or near water, leads to the production of these compounds, effectively mineralizing most organic contaminants \cite{sugama2006glow}. This mechanism of electrolysis is leveraged for the disinfection of water, oxidation of inorganic pollutants, and elimination of undesirable tastes and odors.

Through electrolysis, ozone is also created. Ozone is said to be more effective than the use of chlorine for inactivation of bacteria and viruses, as seen in Table \ref{T1}.  It can be generated with high input energy.

\begin{table}
  \centering
  \caption{Oxidant’s potential}\label{T1}
\begin{tabular}{l|c}
\multicolumn{1}{c}{Oxidant} & Electrochemical Potential (Volts)\\
\hline
Free radical ($•OH$) & 2.8\\
Ozone atom (O) & 2.42\\
Ozone (O3) & 2.07\\
Hydrogen peroxide (H$_2$O$_2$) & 1.78\\
Chlorine dioxide (ClO$_2$) & 1.57\\
Chlorine gas (Cl$_2$) & 1.36\\
Oxygen (O$_2$) & 1.23\\
\hline
\end{tabular}
\end{table}

\section{Experiment}
The experimentation setup consisted of a power supply, reactor and pair of electrodes. The electrodes configuration was point-to-point.

\subsection{Procedure}
The water used for experimentation comes from the stage just before entering the chlorination process in the Puerto Rico’s Aqueduct and Sewer/Sewage Authority (PRASA), and it’s called tertiary water sample from a treatment water plant. PRASA and the PUPR Environmental Laboratory, donated one gallon of tertiary water for this experiment. The water had to be stored in a refrigerated container able to maintain the water’s temperature below 20 \textdegree C. The treatment was done within a period of no longer than 24 hours, before the water go to waste. Likewise, every sample taken, before or after, must be stored at the same temperature, to avoid erroneous results. A small cooler of 3 liters filled with ice was used to store the water during an experimentation process that lasted at least 5 hours. When every component of the system was connected, the initial conductivity, temperature and time was recorded, to begin the treatment process.

\begin{table}[h]
  \centering
  \caption{Average water conductivity}\label{T2}
\begin{tabular}{l|c}
\hline
\multicolumn{1}{c}{Type} & $\mu$S/cm\\
\hline
Distilled water & 0.5 - 3\\
Melted snow & Feb-42\\
Tap water & 50 - 800\\
Potable water in the US & 30 - 1500\\
Freshwater streams & 100 - 2000\\
Industrial wastewater & 10000\\
Seawater & 55000\\
\hline
\end{tabular}
\end{table}

\subsection{Initial conductivity of water}

The electrical conductivity of a water-based solution indicates its ability to conduct electricity, measured as the electrical conductance over a specific distance within that solution. Essentially, conductivity assesses how well water can carry an electrical current, serving as an indicator of the solution's ionic composition. In this project, conductivity measurements were employed to verify if the water quality fell within the desired range for the study's goals. The aim was to evaluate whether plasma discharge could serve as a viable alternative to conventional water purification methods, with a focus on achieving the conductivity levels typical of freshwater streams. Table \ref{T2} displays the average conductivity values for various types of water. \cite{NARS2023}

\subsection{pH of water}
A measure of acidity or alkalinity of water to be tested was done. A scale of 1 to 14 is the value of the potential hydrogen, it was measured 7.80, based on the scale where 7 is a neutral point. The pH value below 7 is acidic, while above 7 is alkaline. The initial pH of the water tested, remained on the neutral point before the reaction. Once the glow discharge was applied to the water, creating ozone oxidation, the pH of water changed to 9.4, meaning that the water pH scale was alkaline, resistant to changes in pH that would make water more acidic. \cite{doi:https://doi.org/10.1002/9781119300762.wsts0002}

\begin{table}
  \centering
  \caption{Experimental conditions of reaction}\label{T3}
\begin{tabular}{l|c}
\multicolumn{1}{c}{Parameter} & Value\\
\hline
Voltage (V) & 850\\
Current (mA) & 450-500\\
Distance between electrodes (mm) & 20\\
Thickness of discharge electrode (mm) & 3.175\\
Initial pH & 7.8\\
Initial temperature (F) & 70\\
Volume (mL) & 400\\
\hline
\end{tabular}
\end{table}

\begin{figure}
  \centering
  \includegraphics[width=.8\columnwidth]{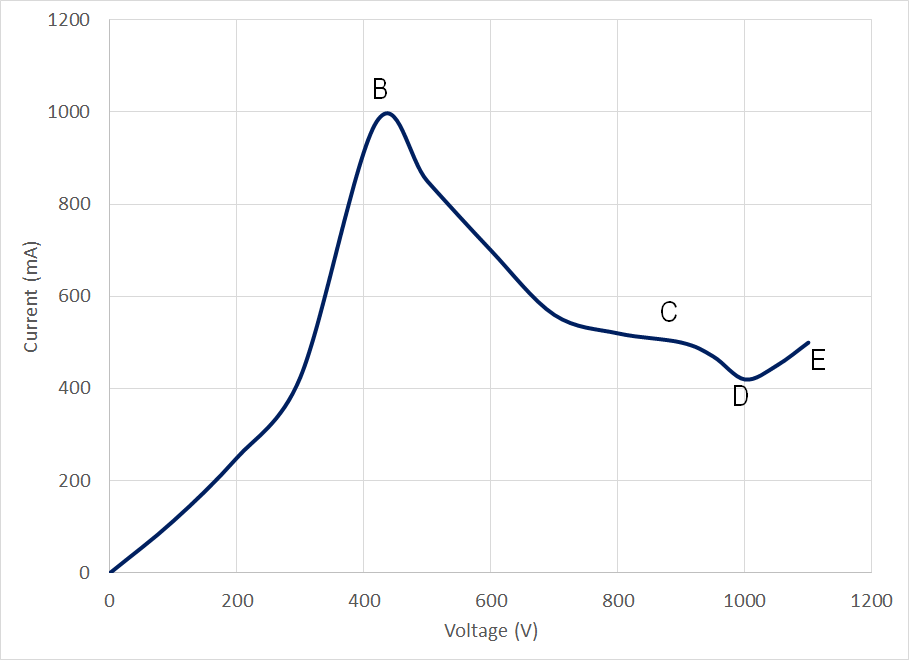}
  \caption{Experimental I-V characteristic}\label{F3}
\end{figure}

When the voltage increased the 550 V with 920 mA, it started to make sparks in the water signaling the breakdown point was reached. The current was on the range of 600 mA to 560 mA and when point C of the I-V characteristic (Figure \ref{F3}) was reached, at a voltage reading of 800V to 850V in the power supply.  At this voltage range, it glow discharge around the anode was observed.

\begin{figure}
  \centering
  \includegraphics[width=.8\columnwidth]{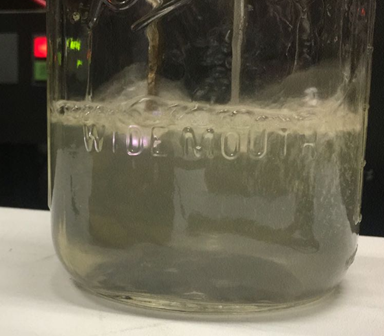}
  \caption{Photograph of the glow discharge at 850V in wastewater.}\label{F4}
\end{figure}

Figure \ref{F3} shows the behavior of Current vs. Voltage throughout the experiment. Water volume and initial temperature may influence the behavior of voltage and current since it can take longer and might need more voltage to reach the breaking point. When the water temperature was increasing, bubbles gas gap were observed around the anode as well as bubbles coming out from it. This means, that when the application of voltage started, it began to create ozone electrolysis on water. When the water reached 190 \textdegree F with a voltage of 800 V, a glow discharge was seen on the anode, as shown in Figure  \ref{F4}.

A magnetic stirrer was used for one of the experiments to evaluate if it helped in keeping the solution mixed, so the water on the bottom would warm up easily and faster. It maintained the water well mixed and warmed up as was expected but the gas gap that needs to be formed around the anode was not uniform, therefore taking longer to produce a discharge from the anode.

Water samples were taken before and after the reaction and placed on Petri dishes. The Petri dishes contained Luria Broth agar to help maintain a focus on culturing E. coli. After 3 days, of observation, it was noted an extensive difference in the Petri dishes containing water sample before the reaction since the Petri dishes had a large percent of bacteria colonies. The Petri dishes with the samples of the water after the reaction did not show bacteria colonies.

The water samples were taken into a specialized external laboratory to make a further study of the presence and absence of E. coli in the water samples. The Most Probable Number (MPN) method \cite{ObliKobu1975, WHOGDWQ84} was used, where a series of tubes were inoculated with dilutions of the water.

The experiment used 15 tubes, 5 were filled with 10 mL of the water treated with the reactor, a the second group of 5 tubes filled with 10 mL of the water treated diluted with distilled water at 10\% concentration.  For last five tubes, 1 mL of treated water was diluted with distilled water, resulting on a mixture of 100 mL, filling the 5 tubes of the third group with 10mL of the mixture.

For this water test, it was used three different dilutions to compare and find bacteria coliforms in one or more of the tubes, this way to validate if the system eliminates bacteria coliforms. The tubes were on observation for 24 hours.

On the first group of 5 tubes, the five tubes resulted negative to the presence of E. coli. The second group of 5 tubes, 3 were positive and 2 negative to the presence of E. coli. The last group of 5 tubes, only 1 resulted positive to E. coli.



\end{document}